# High-count-rate Particle Tracking in Proton and Carbon Radiotherapy with Timepix2 Operated in Ultra-Short Acquisition Time

**C. Oancea,**[a,1] **A. Resch,**[b,] **S. Barna,**[b,c,] **G. Magrin,**[b,] **L. Grevillot,**[b,] **D. Hladik,**[a,]
**L. Marek,** [a,] **J. Jakubek,**[a,] **and C. Granja**[a,]

[a] *ADVACAM*
  U Pergamenky 12, Prague 7, Czech Republic
[b] *MedAustron Ion Therapy Centre*
  Marie-Curie-Strasse 5, 2700 Wiener Neustadt, Austria
[c] *Medical University of Vienna*
  Waehringer Guertel 18-20, 1090 Vienna, Austria

[1]*E-mail*: `Cristina.oancea@advacam.cz`

ABSTRACT: This work investigates the operational acquisition time limits of Timepix3 and Timepix2 detectors operated in frame mode for high-count rate of high deposited energy transfer particles. Measurements were performed using alpha particles from a $^{241}$Am laboratory source and proton and carbon ion beams from a synchrotron accelerator. The particle count rate upper limit is determined by overlapping per-pixel particle signals, identifiable by the hits per pixel counter > 2, indicating the need to decrease acquisition time. On the other hand, the lower limit is the time required to collect the particle deposited charge while maintaining spectral properties. Different acquisition times were evaluated for an AdvaPIX Timepix3 detector (500 µm Silicon sensor) with standard per-pixel DAC settings and a Minipix Timepix2 detector (300 µm Silicon sensor) with standard and customized settings the pulse shaping parameter and threshold. For AdvaPIX Timepix3, spectra remained accurate down to 100 µs frame acquisition time; at 10 µs, loss of collected charge occurred, suggesting either avoiding this acquisition time or applying a correction. Timepix2 allowed acquisition times down to 100 ns for single particle track measurements even for high energy loss, enabled by a new Timepix2 feature delaying shutter closure until full particle charge collection. This work represents the first measurement utilizing Timepix-chips pixel detectors in an accelerator beam of clinical energy and intensity without the need to decrease the beam current. This is made possible by exploiting the short shutter feature in Timepix2 and a customized per-pixel energy calibration of the Timepix2 detector with a larger discharging signal value which allowed for shorter time-over-threshold (ToT) signal. These customized settings extend the operation of the pixel detectors to higher event rates up to $10^9$ particles/cm$^2$/s.

## Contents

## 1. Introduction

The Timepix family of hybrid semiconductor pixelated detectors, including Timepix1 (TPX) [1], Timepix3 (TPX3) [2], and Timepix2 (TPX2) [3], has proven to be a reliable tool for online particle tracking at low particle fluxes. These detectors are used in diverse types of applications, including imaging with X-rays, protons [4-7] and helium [26], radiation treatment and beam monitoring [8, 9], particle tracking [10, 11], particle radiotherapy [12], space radiation monitoring [13], and neutron detection [14, 15]. However, several limitations present in earlier generations, such as Timepix and Timepix3, need to be addressed in the newer Timepix chips (Timepix2 and Timepix4), including their reduced suitability for low-particle fluxes and susceptibility to energy saturation in radiation fields with high linear energy transfer particles (LET). The Timepix2 chip, developed within the Medipix2 collaboration at CERN, represents an advancement over its predecessors, offering new features such as various frame-based operation modes, lower acquisition time settings to cope with high-fluxes of particles, and an adaptive-gain preamplifier to overcome saturation effects [3, 16]. The Minipix SPRINTER, a hybrid semiconductor pixel detector incorporating the Timepix2 chip, was developed by Advacam following the earlier Minipix Timepix3 [17] and Minipix Timepix models. Timepix2 addresses challenges such as track overlap, an issue particularly prominent in radiation therapy environments with particle fluxes exceeding ~$10^5$ particles/cm$^2$/s. These advancements were necessary to overcome limitations in previous versions of the Timepix chips, including the "volcano effect" [18] and the underestimation of deposited energy from highly energetic particles. Timepix2 facilitates the measurement of Time-over-Threshold (ToT) and Time-of-Arrival (ToA) [3], with an architecture designed to prevent incomplete charge collection at the end of the frame by keeping the shutter open until the charge is fully collected.

This study aims to investigate the minimum acquisition time that can be set at the frame level and the impact of varying the per-pixel shaping time parameter, called "Ikrum," on the device's speed. For the first time, the Timepix2 detector has been successfully utilized in measurements within primary beams generated in clinical particle beam settings. Typically, measurements in accelerators require modifications to reduce the flux of particles; previous detectors with Timepix and Timepix3 chips necessitated fluxes less than ~$10^5$ particles/cm$^2$/s. By adjusting the acquisition time in Timepix2 and employing a customized calibration, we successfully imaged single protons and carbon ions at both reduced and clinical fluxes, reaching a maximum of ~$10^9$

particles/cm$^2$/s. To investigate the minimum acquisition time capability of the Timepix2 detector, we conducted experiments with clinical proton and carbon beams, studying its tracking and spectral response under various flux conditions.

## 2. Pixel detectors and experimental setup

### 2.1 AdvaPIX Timepix3 detector

The AdvaPIX Timepix3 detectors [2] used in this study were manufactured by ADVACAM, while the Timepix3 chip was developed by the Medipix Collaboration at CERN. The ASIC read-out chip contains a matrix of 256 × 256 pixels (totaling 65,536 independent channels, with each pixel measuring 55 × 55 μm$^2$) and an active sensor area of 14.08 × 14.08 mm$^2$, resulting in a total sensitive area of 1.98 cm$^2$. The original Timepix also has the same ASIC architecture dimensions. TPX3 provides two signal channels per pixel, which can be set in various modes such as energy and counting, or energy and time of interaction at the pixel level. For this experiment, the detector's operation mode was set to frame mode to simultaneously measure the time of arrival and the deposited energy of individual particles reaching the sensor. The time of arrival can be determined with a resolution of 1.56 ns, while the Time-over-Threshold of the respective pixel, and consequently the deposited energy, can be measured with an energy resolution on the order of several keV [19]. The detector's sensors were manufactured from silicon (Si), with a sensor thickness of 500 μm, and it was calibrated to a minimum threshold of 5 keV. Other Digital to Analog Converter (DAC) parameters were left at their standard values (e.g., Ikrum = 5). The bias voltage was set to 80 V for all measurements. The per-pixel threshold and energy calibration were performed according to Jakubek [20].

### 2.2 Minipix SPRINTER Timepix2 detector

The Timepix2 (TPX2) chip, recently implemented in the Minipix SPRINTER radiation camera by ADVACAM, was utilized in this study. This compact and portable device integrates advanced readout electronics. Equipped with a 300 μm Si sensor, the Minipix TPX2 [21, 22] is smaller than AdvaPIX and has dimensions of 80 mm × 21 mm × 14 mm (see Figure 1a). Power control and data readout were managed through a single USB 2.0 connector with a cable length of up to 3 m. The detector was operated in frame mode, achieving a data frame rate of up to 67 fps (Energy ToT with 14-bit channels) or 99 fps in other operation modes. The Timepix2 chip's two per-pixel signal channels are configured for optimal operation, allowing the registration of events (counting), deposited energy (ToT with 14-bit channels), and time of interaction (time-of-arrival, ToA with 1.56 ns resolution). The ToT clock frequency is set at 10 MHz. The detector was operated using both the standard per-pixel calibration and a customized calibration where the Ikrum was set to 30 (approximately 7.2 nA) [27] and the threshold to 10 keV. The bias voltage was maintained at 80 V for all measurements. The per-pixel threshold [21] and energy calibrations were carried out using the adaptive gain preamplifier enabled. Data were processed using Track Processing and for particle types identification, algorithms based on artificial intelligence (AI) single-layer neural networks (NN) from DPE-TraX Engine [23] were used.

### 2.3 Alpha particles: low-flux

A simple experimental setup was designed in air, placing a $^{241}$Am source (a point source) with energy of 5.5 MeV on top of the detectors AdvaPIX TPX3 and Minipix TPX2, respectively. A schematic representation of the experimental setup is shown in Figure 1a. The radiation source was fixed to its support using glue. Active water cooling was used to maintain a constant detector temperature of 22°C, ensuring stable operation. Data were analyzed across the entire sensor area.

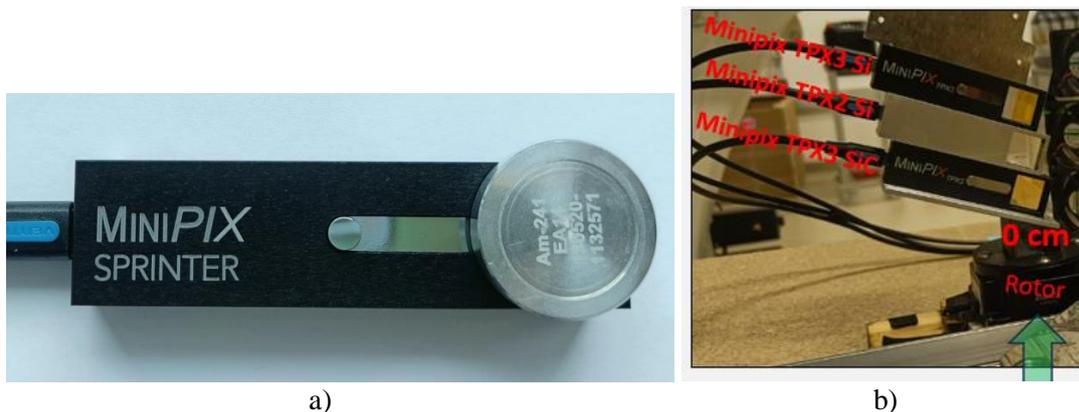

a)  b)

**Figure 1.** a) Schematic illustration of the experimental setup in air using an alpha source, [241]Am, positioned above the silicon sensor of the MiniPIX SPRINTER detector, equipped with a TPX2 chip and a 300 µm thick Si sensor. b) Experimental setup at MedAustron, utilizing proton and carbon beams. Beam direction is marked by the green arrow.

## 2.4 Proton and carbon beams, high flux

MedAustron is a carbon and proton beam therapy facility [24] that accelerates particles to high energies, ranging from 120 to 402 MeV/u for carbon ion beams and from 62 to 250 MeV for proton beams. It can operate at high fluxes, reaching up to $10^9$ particles/cm$^2$/s in clinical mode. For the proton beam, a reduced current mode is also available, where the particle flux is approximately $10^5$ particles/cm$^2$/s. This study aimed to evaluate single particle detection in the clinical fluxes of both proton and carbon-ion beams. Figure 1b shows the experimental setup at MedAustron experimental room. Measurements were carried out in air. The Minipix Timepix2 detector was positioned at the isocenter, with single energy pencil beams delivered directly to the center of the detector.

## 3. Results

## 3.1 Single particle detection in low-intensity particle rate– alpha particles [241]Am

In the experimental setup shown in Figure 1a, using alpha-particles the acquisition time was varied from 1 ms to 10 µs to examine the lower acquisition time limits of the detectors. The particle count rate upper limit is determined by overlapping per-pixel particle signals, identifiable by the hits per pixel counter > 2, indicating the need to decrease acquisition time. On the other hand, the lower limit is the time required to collect the particle deposited charge while maintaining spectral properties. The histogram of deposited energy measured by TPX3 and TPX2 at different acquisition times can be seen in Figure 2. The peak values were compared using the longest acquisition time as the reference spectrum while the acquisition time was gradually decreased. In case of TPX3 detector (Figure 2a) at acquisition time of 0.00001 s (10 µs) the main peak corresponding to alpha particles is missing. In this case, the deposited energy spectra ends before 2400 keV, due to short acquisition time and incomplete charge collection. Thus, when setting a frame acquisition time <100 µs, this effect should be accounted for [25]. Ideally, in case of AdvaPIX TPX3 the recommended lowest acquisition time to be set above ~100 µs if no additional corrections are used. In a similar experimental setup, data were collected at various acquisition time values using the TPX2. As shown in Figure 2b, the energy spectra remained consistent in position with the spectrum obtained at the largest acquisition time for acquisition times as low as 1 us. This consistency aligns with the design of the Timepix2 chip, which in energy mode operation delays the shutter closure until all charge from incoming particles is fully collected.

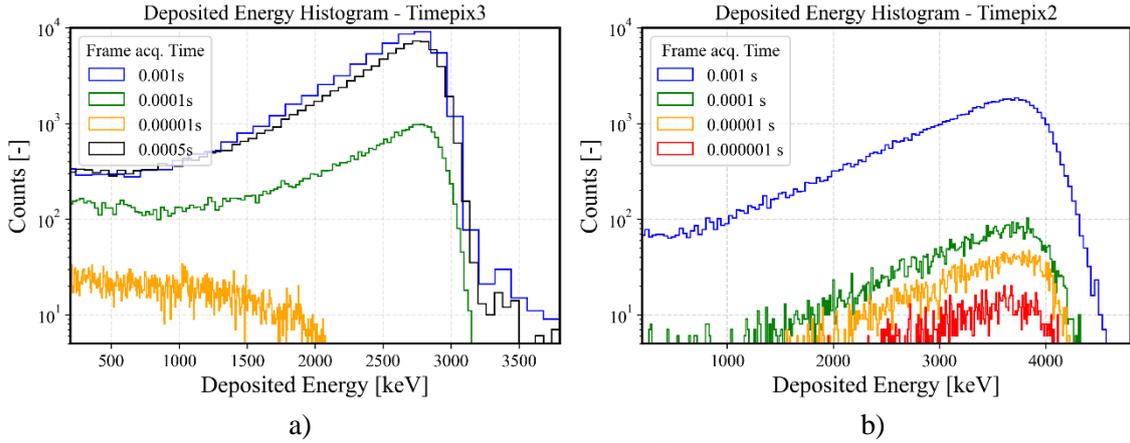

a)                                               b)

**Figure 2**. Deposited energy spectra of alpha particles from $^{241}$Am measured in air for various detector acquisitions times: 1 ms (blue), 100 µs (green) and 10 µs (orange). Data were collected in frame mode with acquisition time a) down to 10 µs (orange) of frame acquisition by the AdvaPIX TPX3 Si detector and b) for the TPX2 Si detector with frame acquisition time down to 1 µs (red).

### 3.2 Visualization of radiation field at 100ns frame acquisition time

Further was investigate if the Minipix SPRINTER Timepix2 detector can operate at even shorter acquisition times, at the nanosecond level, for measurements in primary high-flux proton and carbon beams produced by an accelerator. Figure 3 presents a radiation field visualization for two types of particle beams, as measured by the TPX2 detector with a frame acquisition time of 100 ns. In Figure 3a can be seen the radiation field for 97 MeV protons, while Figure 3b shows the radiation field for a 238.6 MeV/u carbon ion beam. Both visualizations represent the spatial distribution of deposited energy by single particles for a selected region of the sensor of 100 ×100 pixels (5.5 × 5.5 mm$^2$). The visual examination of clusters in both figures indicates complete charge collection for both types of particles, even with the extremely short acquisition time of 100 ns per frame. This demonstrates the detector's capability to accurately capture the energy deposition patterns of individual particles within high-energy beams.

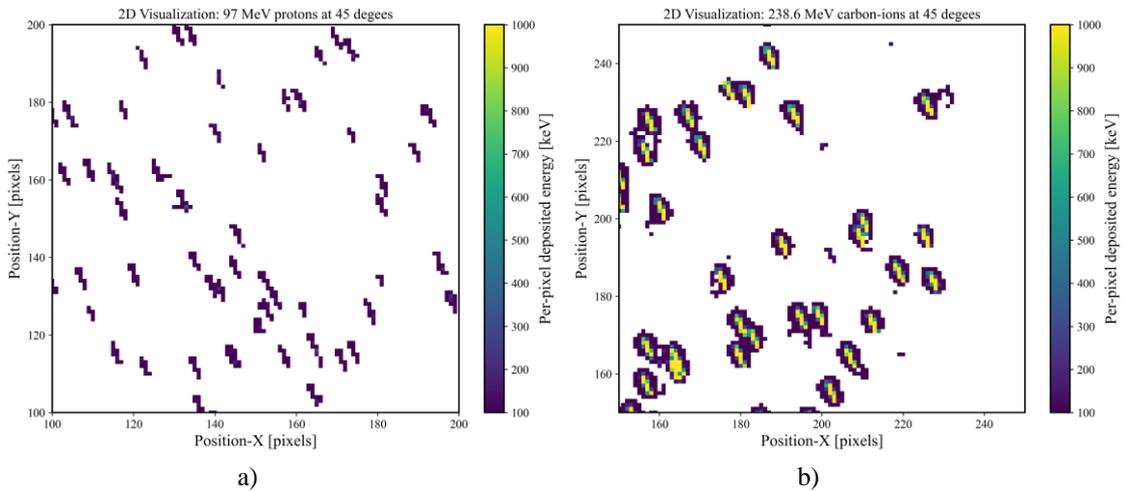

a)                                               b)

**Figure 3**. Visualization of radiation fields showing single particle tracks produced by a) 97 MeV protons and b) 238.6 MeV/u carbon ions, as measured by the TPX2 at a 45-degree incident angle to the sensor plane and with a frame acquisition time of 100 ns. Measurements in proton beams were carried out with standard

calibration settings, while measurements with carbon-ion beams were performed using customized detector DAC parameters.

### 3.3 Energy spectrum of proton and carbon beams delivered at high-fluxes

Using the TPX2 detector, the deposited energy spectra of single particles were measured, as shown in Figure 4a for protons and Figure 4b for carbon ions. These spectra were obtained from the data presented in Figure 3. Figure 4a displays the deposited energy spectra measured using the Timepix2 detector with standard DAC parameters (Ikrum = 5 and threshold = 5 keV) measured in proton beams. This figure illustrates the effect of varying the acquisition time per frame on the deposited energy spectra. The energy spectra exhibit the same shape and the mode at ~550 ± 77 keV regardless of the acquisition time. The corresponding deposited energy calculated from SRIM taking into account the particle 3D trajectory of incoming particles at 45 degrees and the stopping power of 97 MeV protons in Si ($S_{Si}$=1.39 keV/μm) is 569.90 keV (±4%). Figure 4b shows the deposited energy spectra of single particles from a 238.6 MeV/u carbon ion beam, with the sensor tilted at 45 degrees. In case of carbon ion beams the measured energy peak corresponded to ~10.5 ± 0.9 MeV whereas the data from SRIM showed a deposited energy of 10.9 MeV (±4%).

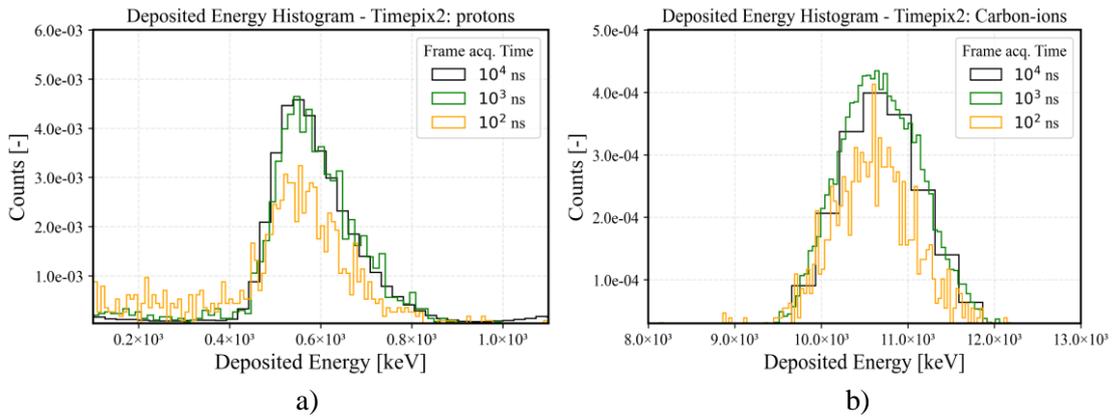

**Figure 4.** Deposited energy spectra of single particles registered by TPX2 at various acquisition times for: a) 97 MeV proton beam with standard detector configuration and b) 238.6 MeV/u carbon beam with customized detector configuration.

### 3.4 Measurements in carbon ion beams at clinical fluxes

The performance of the TPX2 in high-intensity $^{12}C$ particle beams at clinical fluxes up to $10^9$ /cm$^2$/s was evaluated. When imaging the radiation field with standard configuration at these fluxes, there were incomplete tracks with partially collected charge. To address the issue of incomplete tracks, a customized calibration was employed. This involved implementing a higher Ikrum setting of 30 and a threshold (THL) of 10 keV to enhance the detector's response speed. This customized configuration allowed for faster data collection, enabling acquisition times as short as 100 ns for imaging single particles. One limitation to note when using low-acquisition time is the dead time following each frame, which reduces the detector's duty cycle. Despite this, the small duty cycle is manageable and does not impede the utility of the detector, especially in high-count-rate scenarios where continuous single-particle imaging and data collection is not feasible with conventional detectors.

Figure 5 provides understandings into the impact of different calibration methods on radiation field visualization in carbon beams, crucial for optimizing detector performance and accuracy in radiation dosimetry applications. Figure 5a shows the radiation field with the customized calibration, where the full clusters of single particles are clearly visible, indicating complete charge collection. In contrast, Figure 5b shows the radiation field with the standard configuration, where incomplete charge collection results in fragmented or broken tracks.

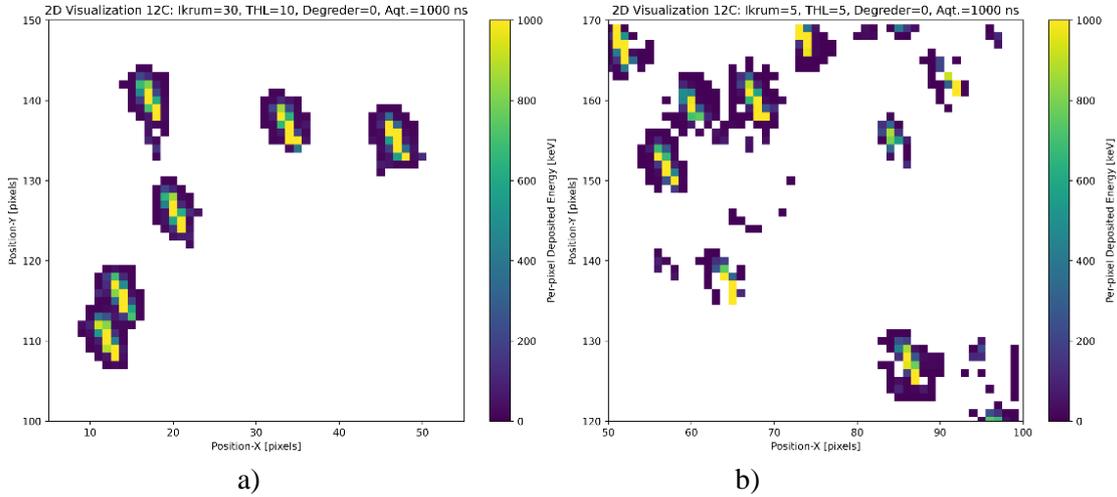

a)                  b)

**Figure 5**. Radiation field visualization of a 238.6 MeV/u carbon beam at clinical intensity for two configurations of the TPX2 at the same acquisition time of 100 ns: a) Customized calibration at Ikrum = 30, THL = 10 keV; b) Standard calibration at Ikrum = 5, THL = 5 keV. Only a portion of the detector pixel-matrix is shown (50 px x 50 px = 2.75 mm x 2.75 mm = 0.756 $cm^2$).

Figure 6 illustrates the particle flux over time, comparing the standard configuration (solid lines) with a customized configuration (dashed lines) for particle types identified using AI NN [23]. Three classes of particles were identified: protons (blue), electrons and photons (orange), and carbon ions (green). The total particle flux is plotted as a function of time.

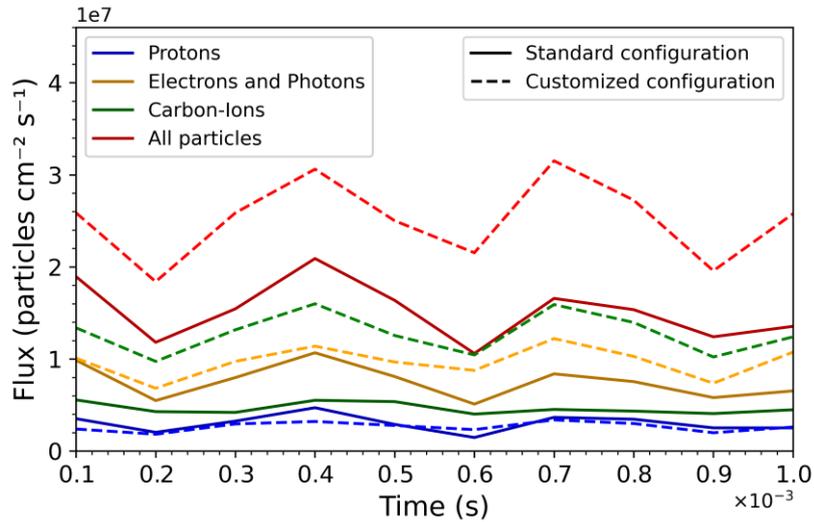

**Figure 6**. Flux of particles over the entire sensor area (~2 $cm^2$), with particle identification (PID) classified into carbon ions, electrons and photons, and protons using AI neural networks. The comparison between standard configuration (solid lines) and customized configuration (dashed lines) highlights the impact of calibration settings on the detection of different particle types.

The flux of protons remains relatively stable across both configurations, with an average ratio (customized/standard) of 0.96, indicating a slightly lower flux in the customized setup. Carbon ions show the most significant difference between configurations, with the customized setup resulting in nearly a threefold increase in flux (ratio of 2.89). For electrons, the customized configuration led to a 40% higher detected flux (ratio of 1.40). It is noteworthy that carbon ions with incomplete charge collection are recognized by the AI NN algorithms as electrons and photons. This misclassification emphasizes the importance of complete charge collection for accurate particle identification.

## 4. Conclusions

The newly developed and calibrated detection system based on Timepix2 chips effectively characterizes radiation fields through single-particle detection and identification.
For the first time, TPX2 was used for single-particle detection in clinical beams with high fluxes, utilizing a customized calibration combined with nanosecond-scale frame acquisition times. This work demonstrates the extension of the operational range of Minipix Timepix2 to accommodate high-flux particle environments while maintaining accurate spectrometry, even for high-energy loss particles, which was not possible for AdvaPIX TPX3. The TPX2, with its customized configuration (Ikrum=30, THL=10 keV), allows for the detection and imaging of individual particles at fluxes up to $10^9$ particles/cm²/s, such as those encountered in clinical proton and carbon-ion therapy beams. The ability to capture particle tracks at acquisition times as short as 100 ns—without loss of spectral accuracy—indicates a considerable improvement in particle tracking performance, particularly for high-energy loss particles that typically require longer pulse shaping times. This capability, especially in the context of radiation therapy applications, offers the potential to measure in clinical mode. This work confirms that the Minipix Timepix2 can operate efficiently in high-flux environments, offering a robust solution for particle tracking and energy deposition measurements, expanding the detector's applicability to fields requiring precise measurements under extreme conditions, such as particle radiotherapy and beam diagnostics.